# Software Change Request in AI-Synthesized Software Evolution

**Muhammad Nadeem[a],[b], Khurram Shahzad[c], Muhammad Nadeem Majeed[c]**

[a] Department of Computer Science, FCIT, University of the Punjab, Lahore, Pakistan; [b] Department of Artificial Intelligence, SST, University of Management and Technology, Lahore, Pakistan; [c] Department of Data Science, FCIT, University of the Punjab, Lahore, Pakistan
muhammad.nadeem@pucit.edu.pk; muhammad.nadeem@umt.edu.pk; khurram@pucit.edu.pk; nadeem.majeed@pucit.edu.pk

## Abstract

Mobile app stores are a rich source of user feedback containing many change requests, bug reports, improvement suggestions, etc., relevant to the post-deployment modification of software. However, it is challenging to manually translate this free format data to actionable change requests (SCRs).

This review presents studies that have applied rule-based techniques, traditional ML approaches, topic modelling, hybrid approaches, and a few deep learning approaches. The data sets used range from small annotated sets to millions of reviews. However, there is inconsistency in the names, availability, annotation guidelines and sampling methods used across different studies. Many studies use Bag-of-Words and TF-IDF, while Naïve Bayes, Support Vector Machines, Decision Trees and Random Forest are frequently applied. Precision, Recall and F1-scores are commonly used for evaluation; however, results and experiments are not easily comparable. Some tools have been developed such as SURF, ARDoc, ChangeAdvisor, ALERTme, AUREA, DIVER, ACRM and KEFE to partially automate the classification, summarization, topic detection and traceability. This review shows that there is still a lack of consistency in the data sets used, reproducible annotation, generalization across languages and domains, contextual modelling, interpretability and integration with maintenance activities. This review provides a consolidated view and taxonomy to provide a research agenda to support more reliable AI enabled software evolution.

## 1. Introduction

A Software Change Request (SCR) is a documented proposal to modify a software product or its behaviour. Software maintenance standards and literature distinguish corrective, adaptive, perfective and preventive maintenance [32, 34]. Change control authorities examine the proposed changes, effects, resource requirements and quality considerations [26].

App store reviews provide an important source of feedback after deployment of mobile applications. The reviews will typically be informal and will describe issues faced by users when using the application such as crash issues, compatibility issues, usability issues, missing features, security issues and suggestions for improvements. Not all reviews will be SCRs. Some may be positive, some may ask questions or be irrelevant. The scale and unstructured nature of this feedback make manual analysis costly and slow. Natural Language Processing (NLP), ML and deep learning (DL) can support automated identification, classification, summarization and prioritization of relevant feedback [1,7, 16, 20, 24, 25].

While existing work has focused on specific tasks and has provided tools for review mining, the dataset, labelling, features and evaluation criteria differ across works. This makes it difficult to reproduce results and compare between different methods. We provide a review of the evidence for SCR extraction from mobile app reviews and cover the entire research process, from data collection, labelling, modelling, evaluation and to producing outputs.

## 1.1 Review objectives and contributions

The objective of this review is to systematically examine AI-supported analysis of mobile application reviews for Software Change Request (SCR) identification and app evolution by investigating the artifacts, classification levels, and taxonomies proposed in existing studies; evaluating the datasets, sampling strategies, and annotation procedures used to support these approaches; analysing the features, algorithms, and evaluation measures employed, along with their reported performance; and identifying existing limitations, research gaps, and future opportunities. The review contributes a consolidated understanding of SCR-related artifacts and classification approaches, a comparative synthesis of data collection and annotation practices, an assessment of AI techniques and their reported effectiveness, and a research agenda to guide the development of more reliable and effective AI-supported mobile application evolution solutions.

## 2. Review Method

### 2.1 Protocol and search strategy

The review follows Kitchenham's three phase process—planning, conducting and reporting which is widely used in software-engineering SLRs [30]. This three phase process also visually presented in Figure 1. The The protocol provided details about the research questions, sources of information searched, keywords, inclusion criteria, quality appraisal and data extraction areas. Ten information sources were searched including; ACM Digital Library, IEEE Xplore, IEEE, Google Scholar, Web of Science, Springer, Wiley, Cambridge Core, Scopus and JSTOR. Search strings combined terms for "software change requests" or "software maintenance," "mobile app reviews," NLP/ML/DL methods, and review content such as features, functionality, updates, issues and problems. Backward and forward citation tracing through Google and Google Scholar supplemented the primary search. The search and selection process is summarized in Table 1.

| Review element | Procedure |
|---|---|
| Publication period and language | English-language studies published 2015–2024. |
| Sources | ACM, IEEE Xplore, IEEE, Google Scholar, Web of Science, Springer, Wiley, Cambridge Core, Scopus and JSTOR. |
| Search concepts | SCR/software maintenance AND mobile app reviews AND NLP/ML/DL/neural methods AND feature, update, functionality, issue or problem terms. |
| Supplementary search | Backward and forward citation tracing; advisory review of coverage. |
| Selection outcome | The article reports staged reductions from approximately 3,500 records to 25 included studies: 3,500→2,000→600→500→200→25. |

**Table 1**

Summary of Search and Selection Process

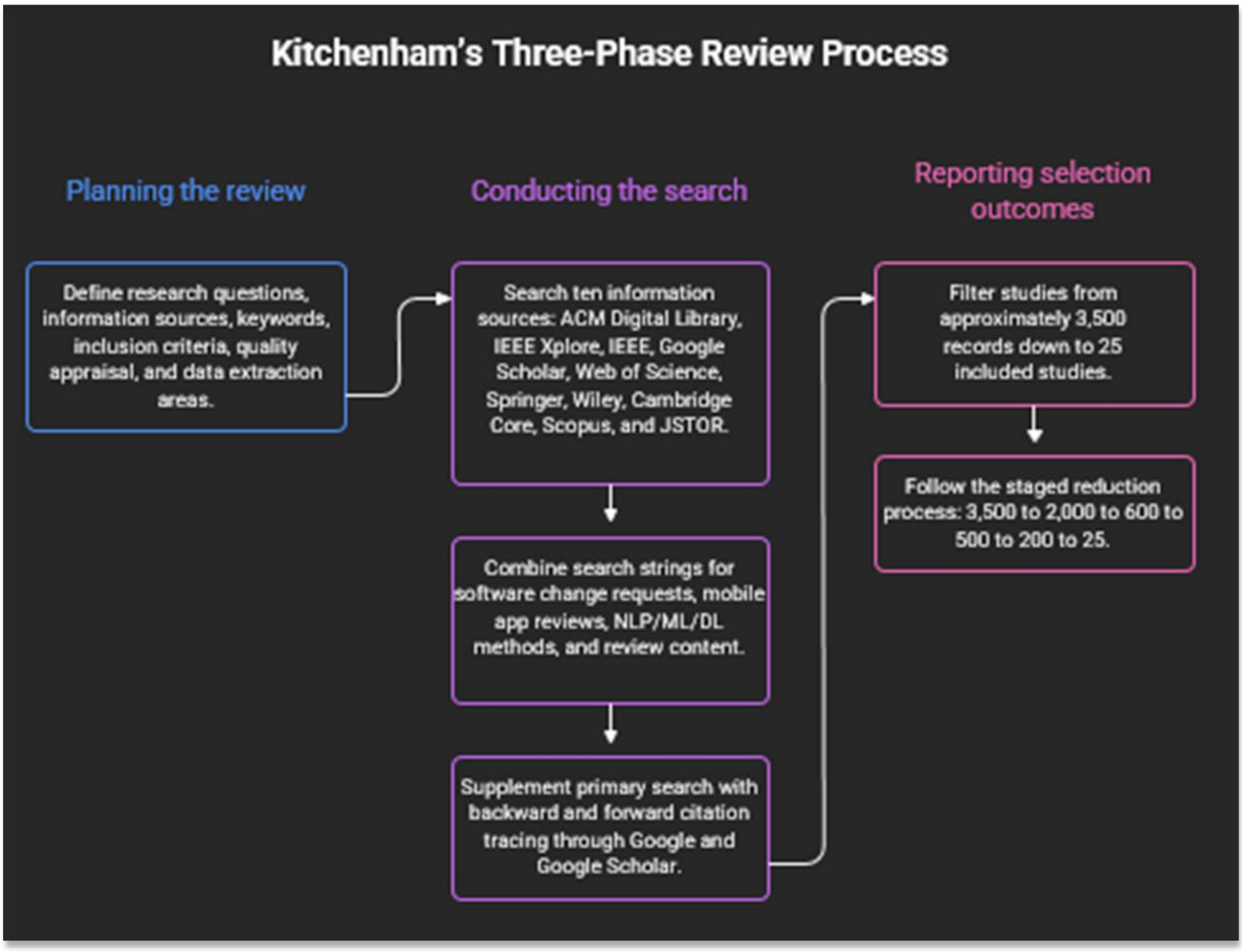

**Figure 1:** Kitchenham's Three-Phase Review Process

## 2.2 Eligibility and screening

Studies that focused on mobile applications, performed analysis of user feedback reviews and used NLP techniques such as rule-based, ML and DL, relating to post-deployment. The studies considered feedback in relation to corrective, adaptive, perfective and preventive maintenance. We excluded studies that involved non-mobile software, pre-deployment requirements engineering, surveys and documentation (not reviews). Other feedback channels included e-mail, interviews, social media, usability testing and website feedback. Feedback was also excluded if only rated or download counts were used. Finally, studies unrelated to project management, design, development, configuration management or quality assurance were excluded.

Screening proceeded in two stages. First, titles and abstracts were assessed against the criteria; second, full texts were examined. A second reviewer rechecked selected records to support consistency. The review describes use of Cohen's kappa to assess agreement, with disagreements discussed and resolved by consensus; a value below 0.6 prompted reassessment. Quality screening was independently conducted by two experts using refined eligibility and quality criteria. The advisory group also reviewed the breadth of the final evidence set. The reported screening sequence should be retained transparently in any replication, including exact record counts at each stage. The pictorial representation of eligibility, screening and data synthesis is given in Figure 2.

## 2.3 Data extraction and synthesis

For each included study, the review extracted publication context, proposed artifact, classification level and type, label scheme, dataset name/source/accessibility, review and sentence counts, app genres, annotation method and annotator count, feature representations, algorithms, evaluation measures, reported scores, contributions and limitations. Findings are presented thematically. As there will be differences between the data and tasks used within each experiment, therefore results should be seen as evidence specific to each study.

# 3. Included Studies and Research Scope

The final corpus contains 25 studies published between 2015 and 2022 in the study table, with the search window extending to 2024. The literature is drawn from several areas including requirements engineering, software maintenance and evolution, software analytics and software quality. Table 2 shows the years that the studies were published. The numbers in brackets indicate their position in the reference list.

| Ref. | Year | Study focus |
|---|---|---|
| [1] | 2015 | Maalej & Nabil : automatic classification of app reviews |
| [2] | 2015 | Yang & Liang : requirements identification and classification |
| [3] | 2015 | Panichella et al. : review classification for maintenance/evolution |
| [4] | 2016 | Di Sorbo et al. : summarizing reviews to recommend changes |
| [5] | 2016 | Panichella et al. : ARdoc review classifier |
| [6] | 2016 | McIlroy et al. : automatic labelling of user issues |
| [7] | 2016 | Maalej et al. : automatic classification of app reviews |
| [8] | 2017 | Grano et al. : Android app/user-feedback dataset |
| [9] | 2017 | Lu & Liang : non-functional requirements classification |
| [10] | 2017 | Palomba et al. : recommending/localizing change requests |
| [11] | 2017 | Guzman et al. : mining tweets for requirements (contextual study) |
| [12] | 2017 | Di Sorbo et al. : SURF review summarizer |
| [13] | 2018 | Li et al. : mobile app evolution analysis from reviews |
| [14] | 2018 | Gao et al. : emerging issue identification |
| [15] | 2018 | Ciurumelea et al. : automated user-review analyser |
| [16] | 2019 | Stanik et al. : multilingual feedback classification |
| [17] | 2019 | Phetrungnapha & Senivongse : review-to-issue-ticket classification |
| [18] | 2019 | Saito & Klyuev : sentence/review classification with Naïve Bayes |
| [19] | 2019 | Gao et al. : emerging issues from WeChat feedback |
| [20] | 2020 | Al-Hawari et al. : reviews classified into maintenance tasks |
| [21] | 2021 | Yang et al. : BERT and topic-model review requirements analysis |
| [22] | 2021 | Wu et al. : identifying key features from reviews |
| [23] | 2021 | Nadeem et al. : extracting SCRs from mobile app reviews |
| [24] | 2022 | Malgaonkar et al. : prioritizing user concerns |
| [25] | 2022 | Khalajzadeh et al. : human-centric mobile-app issues |

**Table 2**
Yearly Studies with the Main Focus

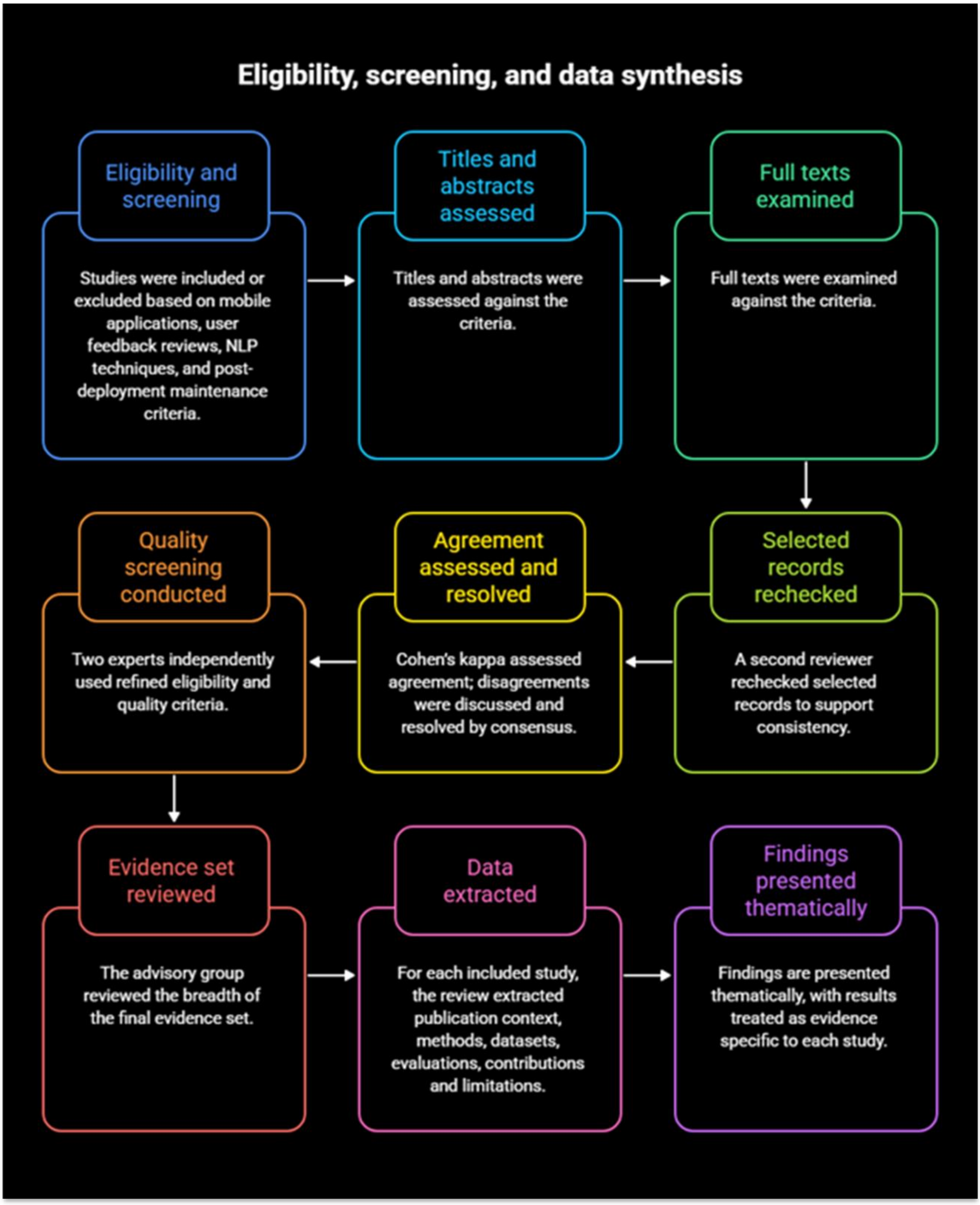

**Figure 2:** Eligibility, Screening and Data Analysis

# 4. Synthesis of Findings

## 4.1 Artifacts and classification approaches

There are three main types of artifacts. Most of the contributions involve proposing a method or analysis. Fewer contributions are a useful tool that can be used. Very few provide re-usable datasets. The methods are either described in text or involve detailed steps and visual pipeline. Visualizing methods helps increase traceability and re-producibility.

The classification approaches vary between very coarse (e.g., SCR or non-SCR, positive or non-positive feedback) and more detailed using multi-class classification (e.g., bug report, feature request, user experience, information seeking (giving, category). Hierarchical and multi-label classifications provide more accurate results when concerns overlap or are nested but need clearer labelling guidelines and more thorough evaluation. Multi-class classification was the most common form of classification found in the corpus, with multi-label classification being much less common. The dimensions, common approaches reviewed with implications are given in the following Table 3.

| **Dimension** | **Common approaches in the reviewed studies** | **Implication** |
|---|---|---|
| Artifact | Methodology, tool, dataset | Most work proposes methods; reusable tools and datasets are less common. |
| Classification level | One, two or three levels | Two-level schemes are common; deeper taxonomies increase granularity and complexity. |
| Task type | Binary, multiclass, limited multilabel/hierarchical | Binary tasks simplify triage; multilabel tasks better represent overlapping user concerns. |
| Labels | Bug, feature request, user experience, ratings; intention categories; functional/non-functional requirements; quality attributes; emerging issues | Label definitions differ, limiting cross-study comparison. |

**Table 3**
Dimensions, Common Approaches and Implications

The reviewed taxonomies include four broad review categories (bug reports, feature requests, user experiences and ratings); intention labels (information seeking, information giving, feature request, problem discovery and other); functional and non-functional requirements; issue-specific labels such as crashes, network/update/compatibility problems, cost, privacy and resource use; and quality attributes such as reliability, usability, portability and performance. Some work also examines app usage, inclusiveness, accessibility, privacy/security and human-centric concerns.

## 4.2 Tools and operational support

The tools address complementary stages rather than providing a complete end-to-end SCR lifecycle. SURF summarizes user-review feedback and supports topic-oriented analysis [4], [12]. ARDoc classifies reviews and supports development-oriented documentation [5]. ChangeAdvisor combines topic modelling, clustering and traceability with issue-tracking integration [10]. ALERTme emphasizes proactive alerts [11]. AUREA combines sentiment and topic analysis [15]. DIVER supports cross-artifact navigation and visualization [19]. ACRM supports clustering and prioritization [20]. KEFE uses knowledge-graph concepts for feature identification and knowledge evolution [22]. AOLDA and the IDEA framework provide analytical/meta-framework support for review analysis [14]. The tools with their main reported roles are given in the Table 4 below:

| **Tool / artifact** | **Main reported role** |
|---|---|
| SURF | Review summarization, sentiment and topic-oriented feedback analysis. |
| ARDoc | Development-oriented review classification/documentation. |
| ChangeAdvisor | Topic modelling, clustering, traceability and issue-tracker integration. |
| ALERTme | Real-time or proactive alerts for improvement requests. |
| AUREA | Combines sentiment and topic analysis to derive actionable feedback. |
| DIVER / WeChat analysis | Emerging-issue exploration and cross-artifact navigation/visualization. |
| ACRM | Clustering and ranking of review-derived concerns. |
| KEFE | Knowledge-graph-based feature identification and knowledge-base evolution. |
| AOLDA / IDEA | Analytical framework for examining review-analysis methods and perspectives. |

**Table 4**
Reported Tools with Main Role

Across these systems, recurring capabilities include review classification, summarization, topic discovery, ranking and recommendation. Differences were apparent between tools in traceability, notification, knowledge representation, sentiment analysis, and integration into development processes. There was no definitive evidence that any of these support a fully automated process from the review to an accepted and verified change.

**4.3 Dataset scale, accessibility and annotation**

Dataset scale is highly variable. The review reports corpora ranging from hundreds or thousands of annotated sentences to millions of app reviews. Examples include datasets with approximately 1.3 million reviews, 1.1 million reviews, 1.14 million reviews and a corpus of about 6.3 million reviews. Sentence-level samples range from roughly 520–1,000 in small exploratory studies to more than one million in larger analyses. The Android data set [8] consists of 395 applications across 23 application genres, with around 288,000 reviews, 450,000+ pieces of feedback and 600 different application versions. The genres range from games, productivity, travel, photography, communication, entertainment, health and fitness, books, business, education, finance, shopping, navigation and tools. Other studies have focused on specific genres, for example books or electronic devices. This makes it difficult to generalize to different types of applications.

Dataset transparency is a major weakness. Only a small subset of studies names and publicly releases a dataset; the review identifies six publicly accessible datasets/resources, while many studies do not provide stable dataset names or links. Most studies collect new data, while a smaller group reuse or combine existing sources. Manual annotation is common, with reported annotator counts from one to twelve; some studies do not report annotator details. Automated annotation is used in a smaller number of studies. The source reports that explicit annotation-guideline creation is almost absent, with only one study identified as having created guidelines. Following Table 5 summarized the dataset/annotation issues, evidence synthesized and research consequence.

| Dataset/annotation issue | Evidence synthesized from the review | Research consequence |
|---|---|---|
| Scale | From small samples (hundreds/thousands) to corpora of millions of reviews. | Scale alone does not ensure representative or comparable samples. |
| Genre | Broad app categories and narrow domain-specific datasets. | Domain imbalance may limit generalization. |
| Availability | Six public datasets/resources identified; many datasets unnamed or inaccessible. | Independent replication and benchmarking are constrained. |
| Annotation | Primarily manual; annotator counts vary from 1–12; some automated pipelines. | Quality, cost and scalability trade-offs remain. |
| Guidelines | Formal annotation guidelines rarely reported. | Label ambiguity and inconsistent annotation reduce reliability. |

**Table 5**
Summary of Dataset/Annotation Issues, Evidence Synthesized ad Research Consequence

For future dataset construction, studies should publish a data statement, collection period, app/version coverage, sampling procedure, label definitions, annotation manual, annotator training, adjudication process and agreement statistics. A combination of expert-reviewed manual labels and carefully validated semi-automatic pre-annotation may balance quality and scale.

### 4.4 Features and modelling techniques

There are six main groups of features: First, basic text features such as Bag-of-Words, TF-IDF, word frequencies and n-grams are often used. These are simple to use however have poor representation of contextual and semantic information. Second, metadata based features such as application categories and release notes are useful for giving context, but are not always available. Third, sentiment and opinion based features are useful for understanding positive/negative feedback but are problematic when dealing with sarcasm and contextual polarity. Fourth, topics and thematic reveal recurring themes and new topics but require interpretation. Fifth, issue-based features link feedback to issues such as bugs, feature requests, compatibility, usability etc. These often require domain knowledge. Finally, linguistic and semantic based features such as POS tags, verb-object relationships and Word2Vec embeddings offer more sophisticated representation but are more computationally and technically complex.

Many of the algorithms used are traditional. Naïve Bayes, Support Vector Machine, Decision Tree, Logistic Regression, J48 and maximum entropy methods were used. Rules based methods and bespoke pipelines using regex, parsing, clustering and domain specific tools were used. Classifier models with topic models, clustering or structured analysis were used. Ensemble methods such as Random Forest, Extra Trees, AdaBoost, Gradient Boosting and XGBoost were used. Topic modelling methods such as LDA and variants were used. Deep-learning adoption is limited in the reviewed corpus; the original synthesis identifies one study using CNN-based classification, alongside later work exploring BERT/topic-model approaches. The article's evidence therefore indicates a transition rather than widespread replacement of classical methods.

### 4.5 Evaluation practices and reported performance

The most commonly presented metrics are precision, recall and F1 score. Accuracy was not as commonly reported, however it can be misleading if there is an imbalance in class distribution. Specific sentiment evaluations were also used. There is a discrepancy between the F1 score within the article. 0.08-1.00 was mentioned in the conclusion whilst 0.57-0.97 was mentioned within the results of experiments. This should be verified against the original papers prior to submitting. Regardless of this discrepancy, the scores are not comparable due to differences in the task, dataset, labelling, class distribution and experimentation. Some papers did not give the full results.

High scores within each study indicate that certain tasks can be learnt well with certain experimental setups. However, this does not show that similar performance can be achieved in other applications. Low or medium scores could be due to small datasets, noisy language, imbalanced classes, poor feature representation or complex task. To fairly compare, the datasets would need to be consistent, with consistent splitting, class specific results and metrics. Future studies should include additional metrics such as Matthews Correlation Coefficient, ROC-AUC/PR-AUC (where applicable), calibration and error analysis and explainability along with precision, recall and F1. Careful consideration should be given to selecting metrics based on the task and class distribution.

### 4.6 Summary of evidence and limitations

The 25 studies establish that app reviews can support automated discovery of defects, feature needs and user concerns. Yet, there are many gaps in the literature. The majority of works use a small number (or none specified) of annotations. Data is often not available and there is a lack of consistency in how labels are categorized. There is also a lack of detail regarding how the annotations were produced. Models are usually specific to domains and languages and therefore results are difficult to compare between studies. Text representations are often shallow and not many studies have used more modern contextual models. Automation is often incomplete. There is also a lack of focus on human factors and multi-label classification. The summary of evidence area with strengths and relevant continual limitations are provide in the following Table 6:

| **Evidence area** | **Strengths** | **Persistent limitations** |
|---|---|---|
| Classification | Established labels and practical ML baselines. | Taxonomies differ; SCR/non-SCR distinction and overlapping labels need stronger treatment. |
| Tools | Summarization, topic discovery, ranking and traceability prototypes. | Fragmented capabilities; limited workflow integration and validation. |
| Datasets | Some large and publicly accessible resources. | Few reusable datasets; uneven genre coverage and incomplete documentation. |
| Models | Classical ML is efficient; hybrid and contextual methods are emerging. | Limited DL evidence; domain/language generalization uncertain. |
| Evaluation | Precision, recall and F1 are widely used. | Inconsistent protocols and incomplete reporting hinder comparison. |

**Table 6**
Summary of Evidence Area, Strengths and Persistent Limitations

## 5. Discussion and Research Agenda

### 5.1 Toward an AI-supported SCR lifecycle

This supports a feedback loop between end users, app stores, analysis pipeline, developers and the organization providing the application. The feedback can be inputted by end-users via app stores. The analysis pipeline will gather and pre-process the reviews, extract relevant SCR information, categories it, determine priority and provide summaries with evidence. Developers then consider the suggested changes. Once changes have been approved, they are implemented and new versions deployed. Further reviews can be given about the release. This cycle is graphically provided in the Figure 3.

In practice, the pipeline should keep track of each extracted SCR, the review it came from, the app ID, version number and if possible, time stamp. The system should also know the difference between what the user requested and recommendations made. Before changes are added to the development backlog, they should be reviewed by someone. This could be integrated with existing issue tracking systems and DevOps/CI-CD systems to help traceability, prioritization and monitoring. However, there is no fully validated end-to-end solution within the literature reviewed.

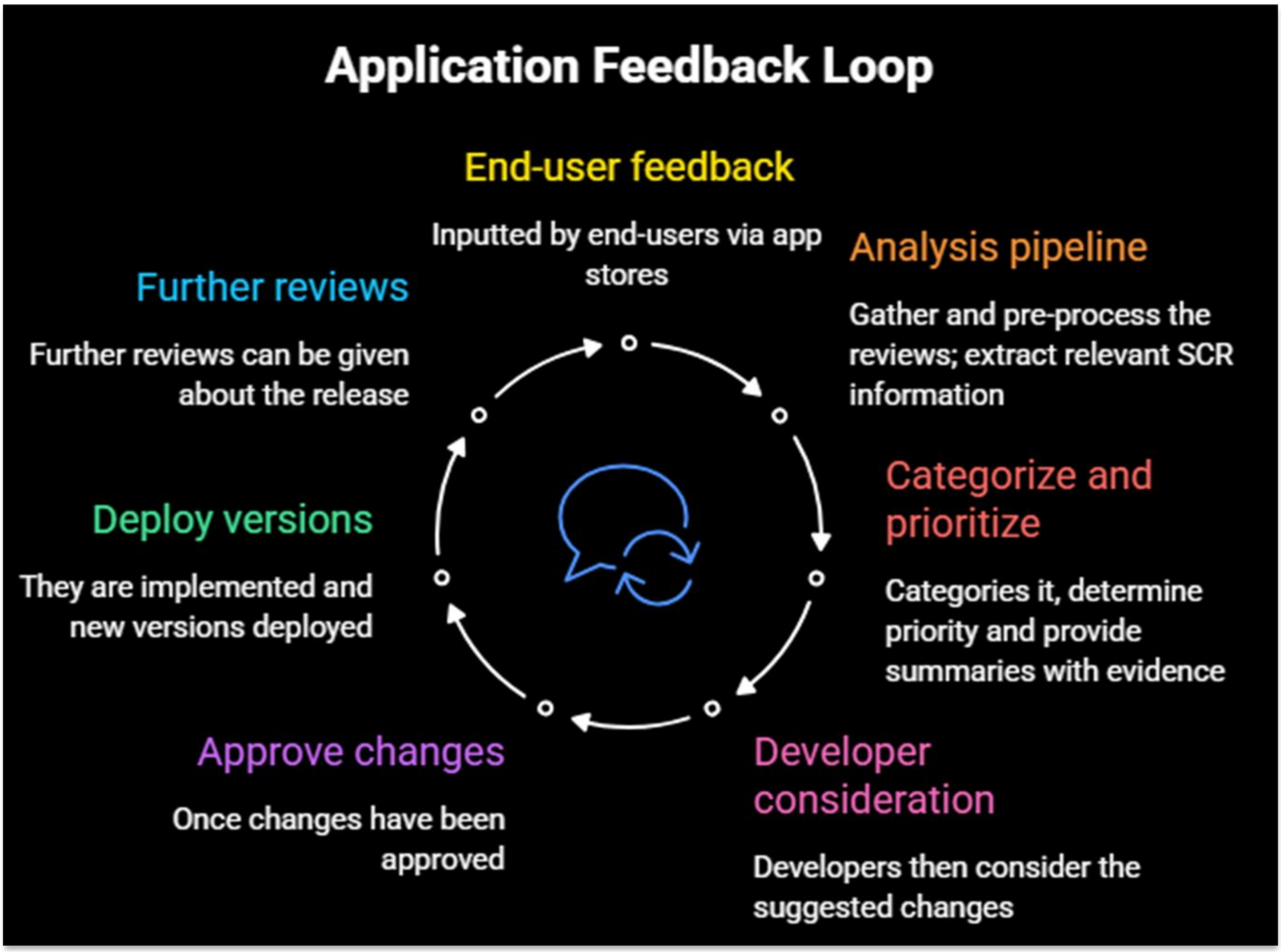

**Figure 3:** App Feedback Loop

### 5.2 Consolidated SCR taxonomy

The reviewed evidence suggests that this can be seen as having four orthogonal dimensions, namely: (i) SCR vs. non SCR; (ii) functional vs. non-functional; (iii) Defect/bug; Feature request; Enhancement; Information giving/seeking; User Experience; Compatibility; Performance; Usability; Privacy/Security; Accessibility and other human related issues. (iv) corrective, adaptive, perfective or preventive. However, this dimensionality shouldn't be flattened by using a single list of labels. Instead it would be better to have a hierarchical or multi-label solution to account for reviews that contain multiple concerns and where issue types are associated with different maintenance activities. These reviewed evidences also pictorially presented in the Figure 4 below:

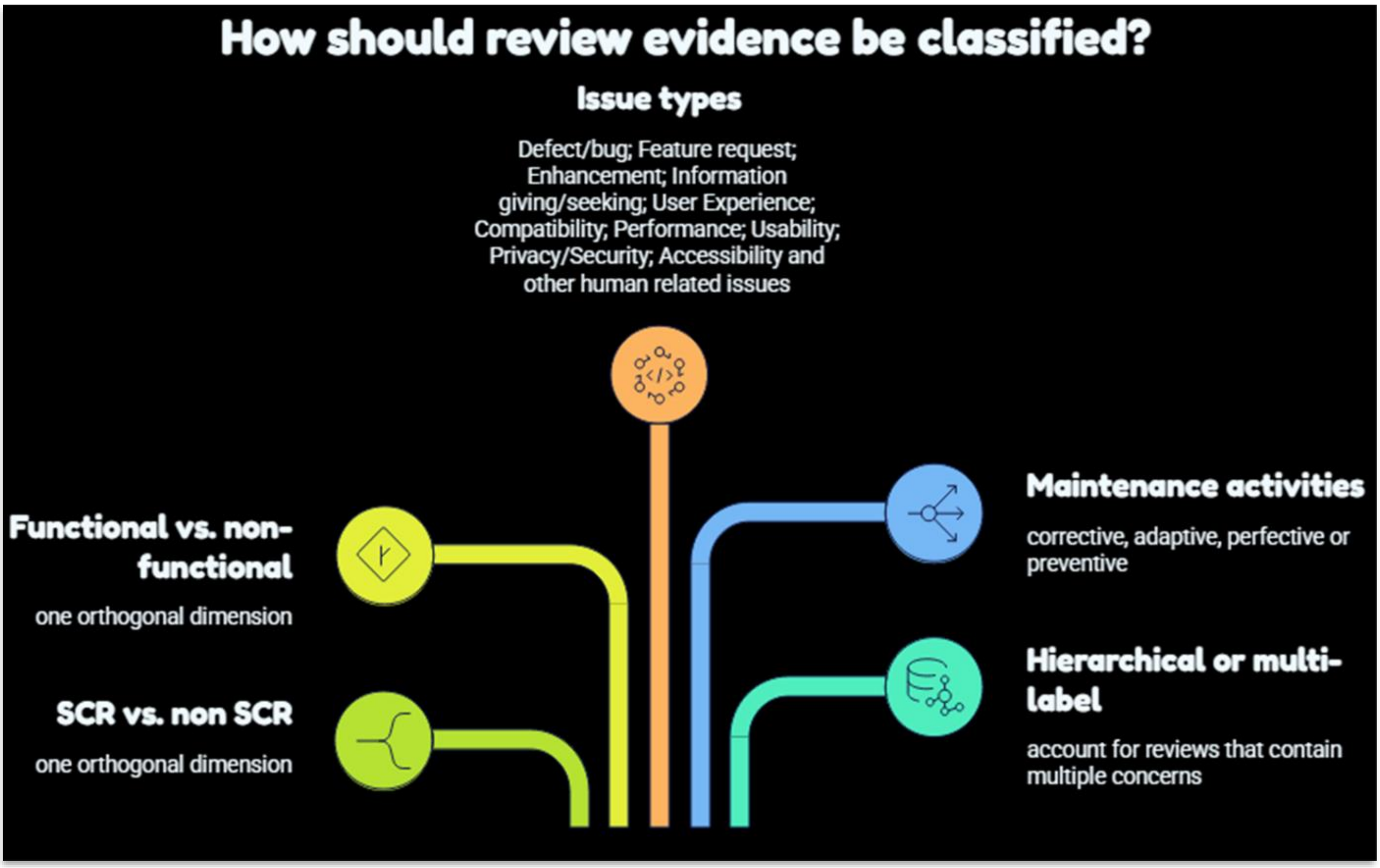

Figure 4: Review Evidence

### 5.3 Priority research directions

The priority research directions as per the underlying review is given in the form of following Table 7 and the Figure 5.

| Priority | Recommended research direction |
|---|---|
| 1. Shared datasets | Construct large, well-versioned, multilingual data sets using consistent identifiers, clearly specified sampling methodology and controlled public access. |
| 2. Annotation standards | Publish label definitions, annotation guidelines, training procedures, judging criteria and inter-annotator agreement. |
| 3. Contextual models | Benchmark using transformer and hybrid models (e.g. BERT and appropriate generative models) against strong classical baselines. |
| 4. Generalization | Demonstrate model transfer capabilities across application categories, platforms, versions, languages and cultures. Provide results on out-of-distribution performance. |
| 5. Robust evaluation | Results are reported using common data splits, per-class precision, recall and F1 scores, MCC and appropriate ROC/PR-AUC values. Model calibration, mistake analysis and confidence intervals provided. |
| 6. Explainability and trust | Provide evidence of spans, review of provenance, uncertainty values and human-in loop checks. |
| 7. Workflow integration | Link review analytic to issue tracking, release management and CI/CD. |
| 8. Privacy and governance | Ensure minimal personal information is collected, in accordance with platform rules and any relevant privacy considerations. Document data retention and access controls. |

**Table 7**
Summary of Priority Research Directions

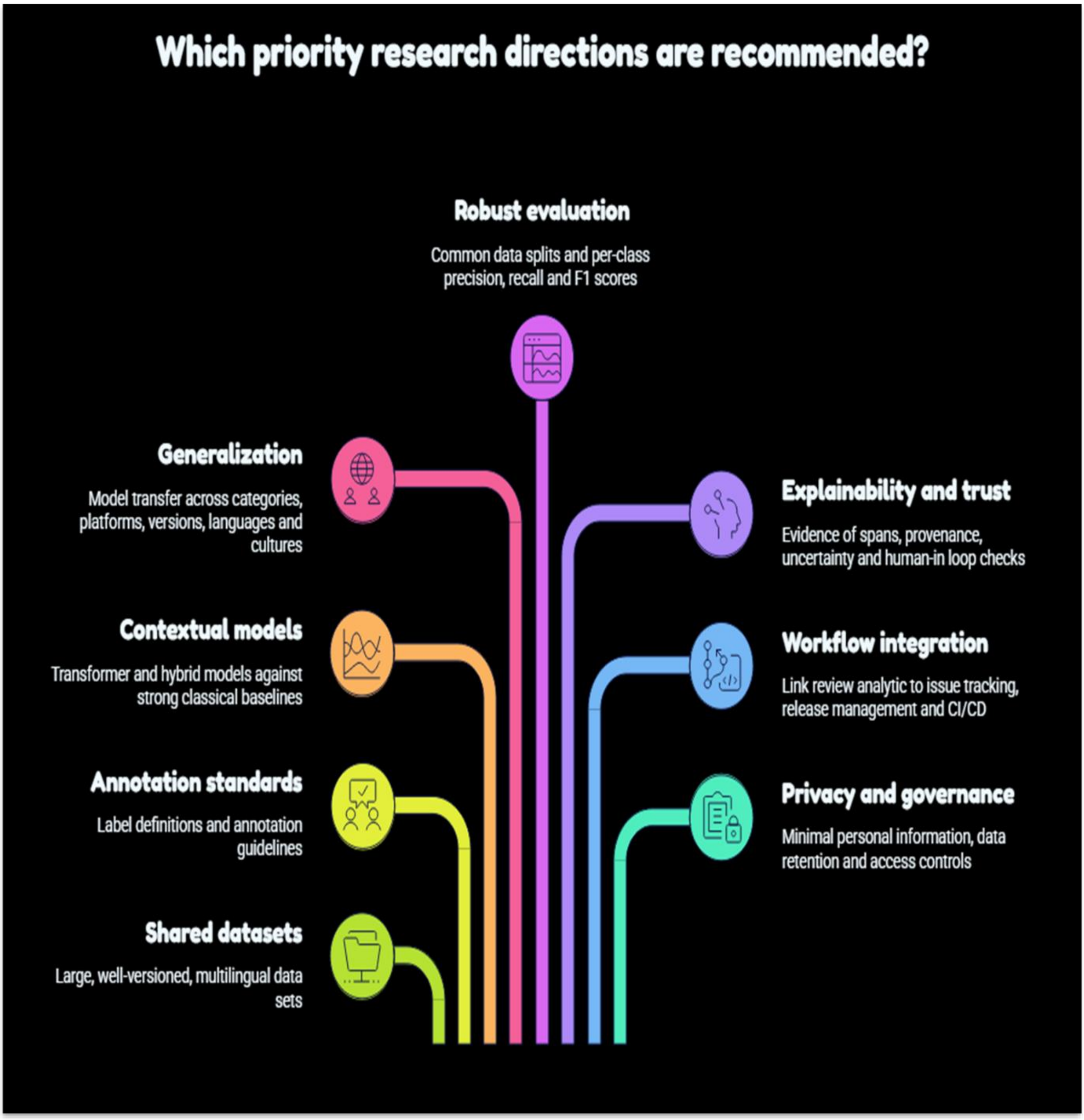

**Figure 5:** Priority Research Directions

### 5.4 Practical implications

This review provides researchers with a foundation for task, dataset and baseline selection. It emphasizes the importance of reproducible methodology. Software development teams can benefit from the automation in the review in terms of reducing manual effort required to identify common issues. However, it is beneficial to build support systems around the classifications. This review can be used by educationalists and students to provide real-world examples of NLP, text classification, topic modelling and software maintenance. Reliable results can be obtained through transparent data handling, clear classifications and by ensuring they have been validated for the application domain.

## 6. Conclusion

This SLR synthesizes 25 studies on extracting and analysing SCR-related information from mobile app reviews. The scientific literature demonstrates that NLP-based, rule-based and traditional ML techniques have long been used to classify bugs, feature requests, user experiences and other issues. Various tools exist to automate certain aspects of the process (e.g., SURF, ARDoc, ChangeAdvisor, ALERTme, AUREA, DIVER, ACRM and KEFE) but comprehensive support is not available. The size of the data varies greatly, little has been made publicly available, whilst annotation guidelines are often not provided. Traditional models are dominant in this area, however there is growing interest using deep learning and transformer-based approaches. This review provides a unified taxonomy of SCR. Furthermore, we identify priorities in developing common datasets, annotation guidelines, multilingual generalization, context modelling, robust evaluation, explainability and DevOps. To make progress in creating AI driven software evolution, we will need better models and also evidence that is transparent, human involvement and end-to-end reproducibility.

### Declaration of AI-assisted tools

ChatGPT and Napkin AI were used to improve language and create some figure graphics during the production of this manuscript. This material has been reviewed, and any edits made. Responsibility for the contents rests with the authors. Please ensure that this sentence is not removed and is updated to reflect which AI tools have been used in the final submission.